\documentclass[draft,11pt,english]{article}
\usepackage{amsfonts,amsmath,amsthm,amscd,amssymb,latexsym}

\newcommand{\supp}{\operatorname{supp}}
\newcommand{\const}{\operatorname{const}}

\numberwithin{equation}{section}

\newtheorem{theorem}{Theorem}%[section]

\begin{document}

\title{
Global in Time Solutions to Kolmogorov-Feller Pseudodifferential
Equations with Small Parameter\thanks{This work was supported by DFG
project 436 RUS 113/895/0-1.}}
\author{S.~Albeverio and V.~G.~Danilov}
%\shorttitle{Short paper title for the headers}
%\shortauthor{F. Author, S. Author}

\date{}

\maketitle

\begin{abstract}
The goal in this paper is to demonstrate a new method for constructing
global-in-time approximate (asymptotic) solutions of (pseudodifferential)
parabolic equations with a small parameter.
We show that, in the leading term, such a solution can be constructed
by using characteristics, more precisely, by using solutions
of the corresponding Hamiltonian system
and without using any integral representation.
For completeness, we also briefly describe the well-known scheme
developed by V.~P.~Maslov for constructing global-in-time solutions.
\end{abstract}

\section*{Introduction}

The goal of the present paper is to present a new approach to the
construction of singular (i.e., containing the Dirac
$\delta$-function as a summand) solutions to the continuity equation
and to show how these solutions can be used to construct the global
in time solution of the Cauchy problem for Kolmogorov--Feller-type
equations with diffusion, potential and jump terms. It is well known
that the asymptotic solutions of the Cauchy problem for linear
equations with a small parameter $\varepsilon>0$ can be constructed
by the WKB method \cite{MF}. In the framework of this method, the
initial partial differential equation is reduced to a system of
equations consisting of the Hamilton--Jacobi equation and several
transport equations. All these equations can be solved under the
assumption that the Hamilton system has smooth solutions
corresponding to the above-mentioned Hamilton--Jacobi equation (the
trajectories of the Hamilton system fiber the phase space). In
general, this ensures only the existence of the classical solution
in small with respect to time. In this case, if, for example, the
Hamilton function is time-independent, then, on the time intervals
where the Hamilton--Jacobi equation has a smooth solution, the
Cauchy problem for this equations (as well as for the corresponding
transport equations) is invertible in time.

It is well known that the above system of equations
(the Hamilton--Jacobi equation and several transport equations)
arises in the construction of WKB solutions of the form
$u_{\text{as}}=\exp\{\frac{i}{h}S(x,t)\}\varphi(x,t)$
for wave equations and Schr\"odinger-type equations \cite{MF}
and in the construction of approximate solutions of the form
$u_{\varepsilon}=\exp\{-\frac{1}{\varepsilon}S(x,t)\}\varphi(x,t)$
for parabolic equations \cite{K1}.

We note that, in the first case (for stationary symbols), the
property of being invertible in time is typical not only for
solutions of this type but also for more general solutions because
of the properties of the equation itself; but in the second case,
the general solutions of the Cauchy problem do not have this
property, and the fact that smooth ``WKB-type'' solutions
$\exp\{-\frac{1}{\varepsilon}S(x,t)\}\varphi(x,t)$ are invertible in
time distinguishes this class of solutions from the other solutions.

Now we explane the time-invertibility condition in more detail and
describe (below in this paper) a method for constructing such
solutions.

First, we note that the function $S(x,t)$ can be defined as
the pointwise limit $S(x,t)=-\lim_{\varepsilon\to0}\ln u_\varepsilon(x,t)$,
where $u_\varepsilon$ is a solution of the equation of parabolic type
with a small parameter \cite{FW1}.

If the function $S(x,t)$ thus defined exists and is smooth, then it
is a solution (classical) of the  Hamilton--Jacobi equation
\cite{FW1,KM1}
\begin{equation}\label{d.1}
S_t+H(S_x,x)=0.
\end{equation}

There is a well-known exact formula expressing the solutions of this equation
in terms of the trajectories of the corresponding Hamilton system
\begin{equation}\label{d.2}
\dot x=H(p,x),\qquad \dot p=-H_x(p,x).
\end{equation}

The fact that the solution of Eq.~\eqref{d.1} is smooth for $t=t_0$
means that the Lagrangian manifold $\Lambda^n_t$ obtained by a shift
of the initial manifold $\Lambda^n_0=\{x,\frac{\partial S}{\partial x}|_{t=0}\}$
along the trajectories of~\eqref{d.2} can be uniquely projected on $R^n_x$
for all $t\in[0,t_0]$.

Since the shift along the trajectories of~\eqref{d.2} is invertible in time,
this implies that the resolving operator of Eq.~\eqref{d.1} is also invertible
in time. Under the same conditions of unique projection,
the solution of the transport equation for the function $\varphi(x,t)$
(the amplitude) is given by the formula
\begin{equation}\label{d.3}
\varphi(x,t)=\frac{C}{\sqrt{J(x,t)}},
\end{equation}
where $C$ is a constant along the projection of the trajectories
of~\eqref{d.2} on $R^n_x$
and $J(x,t)$ is the Jacobian of the mapping of shift along these projections
(the uniqueness of the projection mapping implies the uniqueness and
invertibility of the shift mapping).

Under the above conditions, formulas~\eqref{d.3} are also invertible.
Everything said above can be illustrated by the following simple example.

Let us consider the heat conduction equation
\begin{equation}\label{d.4}
\frac{\partial u}{\partial t}-\varepsilon\frac{\partial^2u}{\partial
x^2}=0,\qquad u\bigg|_{t=0}=e^{-S_0(x)/\varepsilon}\varphi_0,
\end{equation}
where the function $S_0(x)\geq0$ is assumed to be smooth and bounded
together with its derivatives and $\varphi\in C_{0}^{\infty}$, for
example:
$$
S_0(x)=\int^{x}_{0}(1+\tanh z)dz=x+\ln coshx.
$$
Here the following addition condition is satisfied:
\begin{equation}\label{d.5}
\frac{d^2S_0}{dx^2}\geq0.
\end{equation}
The Hamilton function corresponding to \eqref{d.3} is
$$
H(p,x)=p^2
$$
and, respectively, the trajectories of system~\eqref{d.2} have the form
\begin{equation}\label{d.6}
x(x_0,t)=x_0+2t p(x_0,t),\qquad p(x_0,t)\equiv
p(x_0,0)=\frac{dS_0}{dx_0}=1+\tanh x_0.
\end{equation}

Obviously, condition~\eqref{d.5} implies the unique globally-in-$t$ solvability
of the equation
$$
x=x_0+2t\frac{\partial S_0}{\partial x_0}
$$
in i$x_0$ and hence the
global solvability of the corresponding Hamilton-Jacobi equation and
the transport equation in the class of smooth functions.

Further, for Eq.~\eqref{d.3} it is possible to write the Green function,
which is a solution of Eq.~\eqref{d.4},
$$
G(x,\xi,t)=\frac1{\sqrt{2\pi t
\varepsilon}}e^{-(x-\xi)^2/4t\varepsilon},
$$
such that $G(x,\xi,0)=\delta(x-\xi)$. It is well known that one can
write
$$
u=\int G(x,\xi,t) u(\xi,0)\,d\xi.
$$
After the change $t\to-t$, we obtain the Green function for the inverse heat
conduction equation, and it is easy to verify that
\begin{equation}\label{d.7}
u|_{t=0}+O(\varepsilon^N)=\int G(x,\xi,-t)e^{-\hat{S}(\xi,t)/\varepsilon}\hat{\varphi}(\xi,t)\,d\xi,
\end{equation}
where $\hat{S}(x,t)$ and $\hat{\varphi}(x,t)$ are solutions
of the Hamilton--Jacobi equations and transport equation
$$
\frac{\partial\hat{\varphi}}{\partial t}
+2\frac{\partial S}{\partial x}\frac{\partial\hat{\varphi}}{\partial c}
+\frac{\partial^2S}{\partial x^2}\hat{\varphi}=0
$$
at time $t$
satisfying the initial conditions
$$
\hat{S}|_{t=0}=S_0(x),\qquad \hat{\varphi}|_{t=0}=\varphi_0(x),
$$
where $N>0$ is an arbitrary positive number and $O(\varepsilon^N)$
is such that
$$
e^{S_0(x)/\varepsilon}O(\varepsilon^N)  =O(\varepsilon^N).
$$

Equality~\eqref{d.7} can be verified using the Laplace method. The
last equality means that the estimation $O(\varepsilon^N)$ is not
quit suitable here.

There is an absolutely different situation if~\eqref{d.5}
is replaced by the inequality
\begin{equation}\label{d.8}
\frac{d^2S_0}{dx^2}<0
\end{equation}
at least for a certain value of $x_0$, for example,
$$
S_0=\int^{x}_{0}(1-\tanh z)dz=x-\ln coshx .
$$

In this case, the jacobian of the mapping of shift along the
trajectories $x=x_0+2t\frac{dS_0}{dx_0}$ is zero at
$t^*=(\min|2\frac{d^2S_0}{dx^2_0}|)^{-1}$ (in the example, we have
$\frac{d^2S_0}{dx^2_0}=-\cosh^{-2}(x_0)$ and $t^*=1/2$), where the
maximum is taken over all points at which the inequality~\eqref{d.8}
is satisfied. In our example, this is $x_0=0$. For $t>t^*$, the
Lagrangian manifold shifted along the trajectories~\eqref{d.6},
$$
\Lambda^1_t=\{x=x_0-2t(1-\tanh (x_0)),\, p=1-\tanh(x_0)\},
$$
forms an $S$-shaped curve, and there three points of this curve
over a certain point $x$ in the plane $(x,p)$.
These three points are associated with three (local) ``WKB-type'' solutions
$$
u_j=e^{-S_j(x,t)/\varepsilon}\varphi_j(x,t).
$$
It is clear that their linear combination
\begin{equation}\label{d.9}
u=\sum^{3}_{j=1}c_ju_j
\end{equation}
is also a solution, i.e., satisfies the equation with the same accuracy
as each of the functions $u_j$, $j=1,2,3$.
But the functions themselves are not equivalent.

For example, it is clear that the inequality
$$
S_1(\bar{x},t)>S_2(\bar{x},t)
$$
holds at a certain point $\bar{x}$, then the ``WKB'' solutions
$u_1$ and $u_2$ at the point $\bar{x}$ satisfy the relation
\begin{align}\label{d.10}
u_1|_{x=\bar{x}}
&=e^{-S_1(\bar{x},t)/\varepsilon}\varphi_1(\bar{x},t)
\nonumber\\
&=e^{-S_1(\bar{x},t)/\varepsilon}\varphi_2(\bar{x},t)
\big(e^{-(S_1-S_2)/\varepsilon}\varphi_1/\varphi_2)|_{x=\bar{x}}
\nonumber\\
&=u_2 O(\varepsilon^N),
\end{align}
where $N>0$ is an arbitrary number.
This follows from the fact that the difference
$(S_1-S_2)|_{\bar{x}}$ in parentheses in the exponent is positive.

Thus, at each point in formula \eqref{d.9}, it is necessary to choose
the term where the function $S_j$ is minimal.
Such a choice leads to an expression of the form
\begin{equation}\label{d.11}
u=e^{-\phi(x,t)/\varepsilon}\varphi(x,t),
\end{equation}
where $\phi=\phi(x,t)=\min_x\{S_j(x,t)\}$.
It is clear that expression \eqref{d.11} is the leading term
of the approximate solution.
But its substitution into the equation in order to verify that it is
a solution in a certain sense is not a trivial problem,
because the function $\phi(x,t)$ thus determined is not smooth
but is only continuous with a bounded first derivative.

Of course, we can avoid this difficulty if we first calculate terms in \eqref{d.9}
and then pass to \eqref{d.11}.
Such a construction, which takes into account the fact
that the functions $S_j$ and $\varphi_j$
can loose smoothness at the points where the Jacobian
of projection mapping of the Lagrangian manifold on $R^n_x$ is zero,
was proposed by V.~P.~Maslov \cite{M1} and is ideologically similar to
the construction of the Maslov canonical operator
(the Fourier integral operators) \cite{8,10}.
A version of Maslov's construction was proposed in \cite{D5,9}.

The above procedure is quite well for constructing
an asymptotic solution of the Cauchy problem;
some estimates of the discrepancy between the asymptotic and exact solutions
were also obtained.

But attempts to apply these constructions to solve the inverse problem
meet insurmountable difficulties,
and the details of the construction based on the use
of integral representations do not play any role here.

The problem consists in information that we need to obtain
from the solution at $t=T$ and the use to reconstruct the solution
for $0\leq t<T$.

Indeed, as was shown above, if there are singularities
in the projection of the Lagrangian manifold on $R^n_x$,
then the solution can have the form \eqref{d.9} at certain points,
but as was shown above, some of the terms in this sum
are ``infinitely'' small compared with the other.
We can ``measure'' only expressions of the form \eqref{d.11}
which do not contain information about the parts of the Lagrangian manifold
corresponding to the ``infinitely'' small terms of the solution.
But if we move backwards in time, then these (unknown) parts
of the Lagrangian manifold can get into the domains
that is uniquely projected on $R^n_x$, and then they are responsible,
in the projection, for the principal part of the solution.

Thus,
\begin{enumerate}
\item[(i)]
If for $t=T$ the function
$\phi(x,t)=\lim_{\varepsilon\to0}(-\varepsilon\ln u_\varepsilon)$ is
smooth, then the principal part of the solution $u_\varepsilon$ of
the Cauchy problem can be reconstructed for $0\leq t<T$ in the
entire space $R^n_x$.

\item[(ii)]
If for $t=T$ the derivatives of the function $\phi(x,t)$ have singularities
(discontinuities), then, for $0<t<T$,
the principal part of the function $u_\varepsilon$
(the solution of the Cauchy problem)
cannot be reconstructed in the entire domain,
because there are domains where it is impossible to reconstruct the solution.
\end{enumerate}

In our future considerations for non smooth case we will use the
relation between transport and continuity equations. This relation
between the solutions of the continuity equation and the system
consisting of the Hamilton--Jacobi equation plus the transport
equation has been well studied before in the case of a smooth action
functional.

Let the velocity field $u$ be determined as the family of velocities of
points on the projections of the trajectories of the Hamiltonian
system corresponding to the Hamilton--Jacobi equation. In this
velocity field, as it was mentioned by E.~Madelung [13], the squared
solution of the transport equation satisfies the continuity equation
\begin{equation}\label{0.1}
\rho_t + (\nabla, u\rho)+a\rho=0
\end{equation}
with some additional term $a\rho$, which is defined below ($a$ is
equal to 0 if the Hamiltonian is formally self-adjoint). The main
obstacle to the extension of this correspondence globally in time is
the fact that in general the solution of the Hamilton-Jacobi
equation  are smooth only locally in time. The loss of smoothness is
equivalent to the appearance of singularities of the velocity field
mentioned above. Till the recent time there was no method for
constructing formulas for solutions of continuity equation for a
discontinuous velocity field. In Madelung's approach the divergent
form of the continuity equation (in difference with transport
equation) is very important property allows precisely to introduce a
concept of global solution in spite of singularities in the velocity
field.

In the present work we generalize Madelung's approach to the case in
which the singular support of the velocity field is a stratified
manifold transversal to the velocity field trajectories. This holds,
for example, in the one-dimensional case under the condition that,
for any $t\in[0,T]$, the singular support is a discrete set without
limit points.

The class of solutions constructed in this way admits the motion
forward and backward in time. Here we discuss only the construction
as itself, the invertibility problem we plan to discuss in our next
paper.

\section{Generalized solutions\\ of the continuity equation}

Here we follow the approach developed in [1], where the solution is
understood in the sense of an integral identity, which, in turn,
follows from the fact that relation (\ref{0.1}) can be understood in
the sense of distributional space
$\mathcal{D}(\mathbb{R}^{n+1}_{x,t})$. The first step in this way
has been done in [14], see also [15], [16], where the approach based on smooth
approximations of the solutions was used.

We specially note that the integral identities in [1] can be derived
without using the construction of nonconservative products [2, 4] of
the nonsmooth and generalized functions (or measure solutions [5]),
and the value of the velocity on the discontinuity lines (surfaces)
is not given a priori but is calculated. In the case considered in
[1], the integral identities exactly coincide in form with the
identities derived using the construction of a nonconservative
product (measure solutions) in the situation described at the end of
above introduction, which we shall now make more precise.

First, we consider an $n-1$-dimensional surface
$\gamma_t$ moving in $\mathbb{R}^n_x$,
which is determined by the equation
$$
\gamma_t=\{x;t=\psi(x)\},
$$
where $\psi\in C^1(\mathbb{R}^n)$, and $\nabla\psi\ne0$ in the
domain in $\mathbb{R}^n_x$ where we work.

This is equivalent to determining a surface
by an equation of the form
$$
S(x,t)=0
$$
($S\in C^1$ in both variables, $S(x,t)=0$, $\nabla_{x,t}
S|_{S=0}\ne0$) under the condition that
$$
\frac{\partial S}{\partial t}\ne0.
$$

We remain that the situation with $\frac{\partial S}{\partial t}=0$,
can also be covered by making the change of variables
$x'_i=x_i-c_it$ with appropriately chosen $c_i$, $i=1...n$, solving
the problem with the moving surface  and then returning to the
original variables. Possible generalizations are considered later in
this section.

Next, we assume that $\zeta$ belongs to
$C^\infty_0(\mathbb{R}^n\times\mathbb{R}^1_+)$. Then, by definition,
$$
\langle\delta(t-\psi(x)),\zeta(x,t)\rangle
=\int_{\mathbb{R}^n}\zeta(x,\psi(x))\,dx,
$$
where $\delta$ is the Dirac delta function and $\langle,\rangle$ is
the distributional pairing (with respect to the variable
$t\in\mathbb{R}^1_+$ and $x\in\mathbb{R}^n$).

Let $\delta(t-\psi(x))$ be applied to the test function $\eta\in
C^\infty_0(\mathbb{R}^n)$, then
$$
\langle\delta(t-\psi(x)),\eta(x)\rangle =\int_{\gamma_t}
\eta\,d\omega_\psi,
$$
where $d\omega$ is the Leray form [6] on the surface
$\{t=\psi(x)\}$
such that $-d\psi d\omega_\psi=dx_1\dots dx_n$.

One can show that (see [1], [6])
$$
\langle\delta(t-\psi(x)),\eta(x)\rangle =\int_{\gamma_t}
\frac{\eta(x)}{|\nabla\psi|}\,d\sigma.
$$

First, we assume that  the solution $\rho$
to Eq.~(\ref{0.1}) has the form
\begin{equation}\label{2.1}
\rho(x,t)=R(x,t)+e(x)\delta(t-\psi(x)),
\end{equation}
where $R(x,t)$ is a piecewise smooth function
with possible discontinuity at $\{t=\psi(x)\}$:
$$
R=R_0(x,t)+H(t-\psi(x))R_1(x,t),
$$
$e\in C(\mathbb{R}^n)$ and has a compact support, $\psi\in C^2$ and
$\nabla \psi\ne0$ for $x\in\supp e$, and $H(z)$ is the Heaviside
function.

It is clear that the term
$$
e(x)\delta'(t-\psi(x))
$$
appears in (\ref{0.1}) if we differentiate the distribution
$\delta(t-\psi(x))$ with respect to $t$. Hence it is necessary to
have in (\ref{0.1})
$$
( \nabla,\rho u)=-e(x)\delta'(t-\psi) +  \text {smoother summands},
$$
since $\nabla\delta(t-\psi)=-\nabla\psi\delta'(t-\psi)$.
Then we must have
$$
\rho u=\frac{e\nabla\psi}{|\nabla\psi|^2}\delta(t-\psi) +\text{
smoother summands}.
$$

Now we formulate an integral identity, defining a generalized
solution to the continuity equation.

We set $\Gamma=\{(x,t); t=\psi(x)\}$; this is an $n$-dimensional
surface in $\mathbb{R}^n\times\mathbb{R}^1_+$. Let
$$
u(x,t)=u_0(x,t)+H(t-\psi)u_1(x,t),
$$
where $\psi$ is the same function as before, and $u_0,u_1\in C
(\mathbb{R}^n\times \mathbb{R}^1_+)$.

Let us consider Eq.~(\ref{0.1}) in the sense of distributions.
For all $\zeta(x,t)\in C^\infty_0(\mathbb{R}^n\times\mathbb{R}^1_+)$,
$\zeta(x,0)=0$, we have
$$
\Big\langle\frac{\partial \rho}{\partial t} +(\nabla,\rho
u),\zeta\Big\rangle =-\langle\rho,\zeta_t\rangle-\langle\rho
u,\nabla\zeta\rangle.
$$
Substituting the singular terms for $\rho$ and $\rho u$ calculated above,
we come to the following definition.
\medskip

{\bf Definition 1.1}
A function
$\rho(x,t)$ determined by relation (\ref{2.1}) is called a generalized
$\delta$-shock wave type solution to (\ref{0.1})
on the surface $\{t=\psi(x)\}$ if the integral identity holds
\begin{align}\label{1.4}
&\int^\infty_0\int_{\mathbb{R}^n}
(R\zeta_t+(uR,\nabla \zeta)+aR\zeta)\,dx\,dt
\nonumber\\
&\qquad
+\int_{\Gamma}\frac{e}{|\nabla\psi|}\frac{d}{dn_\perp}\zeta(x,t)\,dx=0
\end{align}
for all test functions $\zeta(x,t)\in \mathcal{D}(\mathbb{R}^n\times
\mathbb{R}^1_+)$, $\zeta (x,0)=0$,
$\frac{d}{dn_\perp}=\big(\frac{\nabla\psi}{|\nabla\psi|},\nabla\big)
+|\nabla\psi|\frac{\partial}{\partial t}$.
\medskip

We have also the relation
$$
\int_{\mathbb{R}^n}\frac{e}{|\nabla\psi|}\frac{d}{dn}\zeta(x,\psi)\,dx
=\int_{\Gamma}\frac{e}{|\nabla\psi|}\frac{d}{dn_\perp}\zeta(x,t)\,dx.
$$

We note that the vector $n_\perp$ is orthogonal to the vector
$(\nabla\psi,-1)$, which is the normal to the surface $\Gamma$,
i.e., $\frac{d}{dn_\perp}$ lies in the plane tangent to $\Gamma$.

We can give a geometric definition of the field $\frac{d}{dn_\perp}$.
The trajectories of this vector field are curves lying on the surface
$\Gamma$, and they are orthogonal to all sections of this surface
produced by the planes $t=\operatorname{const}$.
Furthermore, it is clear that
the expression $\frac1{|\nabla\psi|}$ is an absolute value
of the normal velocity of a point on $\gamma_t$, i.e.,
on the cross-section of $\Gamma$ by the plane
$t=\operatorname{const}$,
and the expression
$\frac{1}{|\nabla\psi|}\cdot\frac{\nabla\psi}{|\nabla\psi|}
\overset{\text{def}}{=}\vec V_n$
is the vector of normal velocity of a point on $\gamma_t$.
Thus,
we have another representation:
$$
\int_{\Gamma}\frac{e}{|\nabla\psi|}\frac{d}{dn_\perp}\zeta(x,t)\,dx
=\int_{\Gamma}e\Big((\vec{V}_n,\nabla)+\frac{\partial}{\partial t}\Big)\zeta(x,t)\,dx,
$$
where $V_n=\pi^*(v_n)$, $v_n$ is the normal velocity of a point on
$\gamma_t$, and $\pi^*$ is induced by the projection mapping
$\pi:\Gamma\to R^n_x $.

It follows from the latter definition that the following relations
must hold:
\begin{equation}\label{1.12}
R_t+(\nabla, Ru)+aR\zeta=0, \qquad\text{for all points}\quad (x,t)\not\in\Gamma,
$$
$$
([R]-|\nabla\psi|[Ru_n])+\bigg(\frac{d}{dn}\bigg)^*\frac{e}{|\nabla\psi|}=0,\qquad
\text{for all points}\quad(x,t)\in\Gamma,
\end{equation}

The last relation can be rewritten in the form
\begin{equation}\label{1.3}
\mathcal{K} E+\frac{d}{dn} E =[Ru_n]|\nabla\psi|-[R],
\end{equation}
where $E=e/|\nabla\psi|$, the factor
$\mathcal{K}=(\nabla,\frac{\nabla\psi}{|\nabla\psi|})=\operatorname{div}
\nu$ ($\nu$ is the normal on the surface $\{t=\psi(x)\}$) and, as is
known, is the mean curvature of the cross-section of the surface
$\Gamma$ by the plane $t=\const$,
$\frac{d}{dn}=(\frac{\nabla\psi}{|\nabla\psi|},\nabla)$.

Now we assume that there are two surfaces
$$
\Gamma_i=\{(x,t); t=\psi_i(x)\}
$$
in $\mathbb{R}^n\times\mathbb{R}^1_+$, $i=1,2$,
whose intersection is a smooth surface
$$
\hat\gamma=\{(x,t);(t=\psi_1)\cap(t=\psi_2)\}
$$
belonging to the third surface
$\Gamma_{(3)}=\{(x,t);t=\psi_3(x)\}$. Further, we assume that the
surface $\Gamma_{(3)}$ is a continuation of the surfaces
$\Gamma^{(i)}$ in the following sense. We let $n^{(i)}_\perp$ denote
the curves on the surfaces $\Gamma_i$ and we assign that each
point $(\hat{x},\hat{t})$ on the surface  $\hat{\gamma}$ is assigned
the graph consisting of the trajectories $n^{(1)}_\perp$ and
$n^{(2)}_\perp$ entering $(\hat{x},\hat{t})$ and the trajectory
$n^{(3)}_\perp$ leaving this point (i.e., the trajectories
$n^{(i)}_\perp$ fiber the surface $\Gamma^{(i)}$). We also assume
that the surface (stratified manifold)
$\Gamma_\cup=\Gamma_{(1)}\cup\Gamma_{(2)}\cup\Gamma_{(3)}$ consists
of points belonging to these graphs. Next, we assume that $u(x,t)$
is a piecewise smooth vector field whose trajectories enter
$\Gamma_\cup$.

{\bf Definition 1.2.} Let
$$
u(x,t)=u_0(x,t)+\sum^3_{i=1}H(t-\psi_i)u_{1i}(x,t),
$$
where $\psi$ is the same function as before, and $u_0,u_{1i}\in
C(\mathbb{R}^n\times \mathbb{R}^1_+)$. The function $\rho(x,t)$
determined by the relation
$$
\rho(x,t)=R(x,t)+\sum^{3}_{i=1}e_i(x)\delta(t-\psi_i(x)),
$$
where $R(x,t)\in C^1(\mathbb{R}^n\times\mathbb{R}^1_+)
\setminus\{\bigcup\Gamma^{(i)}_t\}$,
is called a generalized $\delta$-shock wave type solution
to (\ref{1.4}) corresponding to the stratified manifold
$\Gamma_\cup$ if the integral identity
\begin{align}\label{2.4}
&\int^\infty_0\int_{\mathbb{R}^n} (R\zeta_t+(uR,\nabla
\zeta)+aR\zeta)\,dx\,dt
\nonumber\\
&\qquad +\sum^3_{i=1}
\int_{\Gamma_{(i)}}\frac{e_i}{|\nabla\psi_i|}\frac{d}{dn^{(i)}_\perp}\zeta(x,t)\,dx=0
\end{align}
holds for all test functions
$\zeta(x,t)\in \mathcal{D}(\mathbb{R}^n\times
\mathbb{R}^1_+)$, $\zeta(x,0)=0$,
$\frac{d}{dn^{(i)}_\perp}=\big(\frac{\nabla\psi_i}{|\nabla\psi_i|},\nabla\big)
+|\nabla\psi_i|\frac{\partial}{\partial t}$.

As above this relation implies the first equation from (\ref{1.12})
outside $\Gamma_\cup$, equations of the type of the second equation in (\ref{1.12})
on strata $\Gamma_{(i)}$ and the Kirchhoff type relation on $\hat\gamma$:
\begin{equation}\label{1.13}
(e_1+e_2)\big|_{\hat\gamma} =e_{3} \big|_{\hat\gamma}.
\end{equation}

Now we consider the case with $\operatorname{codim} \Gamma>1$.
First,
we note that the second integral in (1.2) can be written as
$$
\int_{\Gamma
}\frac{e}{|\nabla\psi|}\frac{d}{dn_\perp}\zeta(x,t)\,dx
=\int_{\Gamma}e\bigg(\bigg(\frac{\nabla\psi}{|\nabla\psi|^2},\nabla\bigg)
+\frac{\partial}{\partial t}\bigg)\zeta(x,t)\,dx.
$$
We note that if the surface $\Gamma$ is determined by the equation
$S(x,t)=0$ rather than by the simpler equation $\{t=\psi(x)\}$
presented at the beginning of this section, then
$$
\vec V_n=-\frac{S_t}{|\nabla S|}\cdot \frac{\nabla S}{|\nabla S|}
=-\frac{S_t}{|\nabla S|^2}\nabla S
$$
and, of course, the new vector field $\frac{d}{dn_\perp}=(\vec
V_n,\nabla) +\frac{\partial }{\partial t}$ remains tangent to
$\Gamma$.

Therefore, in this more general case, using this new vector $\vec V_n$,
we can again rewrite the integral identity from Definition~1.1 as
\begin{equation}\label{2.5}
\int^\infty_0\int_{\mathbb{R}^n}
\big(R\zeta_t+(uR,\nabla\zeta)+aR\zeta\big)\,dx dt +
\int_{\Gamma}e\Big((\vec V_n,\nabla)
+\frac{\partial}{\partial t}\Big)\zeta(x,t)\,dx =0.
\end{equation}

This form of integral identity can easily be generalized to the case
in which $\Gamma$ is a smooth surface in $\mathbb{R}^{n+1}$ of
codimension~$>1$.

In this case, instead of $\vec V_n$, we can use a vector $\vec v$
that is transversal to $\Gamma$ and such that the field $(\vec
v,\nabla)+\frac{\partial}{\partial t}$ is tangent to $\Gamma$. We
note that the vector $\vec v$ is uniquely determined by this
condition,  which can be treated as ``the calculation of the
velocity value on the discontinuity'' from the viewpoint of [5] and
[7].

Moreover, in this case, the expression for $\rho$ does not contain
the Heaviside function, and it is assumed that the trajectories of
the field $u$ are smooth, nonsingular outside $\Gamma$, and
transversal to $\Gamma$ at each point of $\Gamma$.
In this case,
the function $\rho$ has the form
$$
\rho(x,t)=R(x,t)+e(x)\delta(\Gamma),
$$
where $R\in C^1(\mathbb{R}^{n+1}\setminus\Gamma)$, $e\in
C^1(\Gamma)$, and the  function $\delta(\Gamma)$ is determined
by
$$
\langle\delta(\Gamma),\zeta(x,t)\rangle=\int_{\Gamma}\zeta\omega,
$$
where $\omega$ is the Leray form on $\Gamma$. If
$\Gamma=\{S_1(x,t)=0\cap\dots\cap S_k(x,t)=0\}$, $k\in[1,n]$, then
$\omega$ is determined by the relation, see [6], p.~274,
$$
dt\,dx_1\cdot\dots\cdot d\,x_n = dS_1\cdot\dots\cdot dS_k\omega.
$$

In this case, we assume that the functions $S_k$ are sufficiently
smooth (for example, $C^2(\mathbb{R}^n\times\mathbb{R}^1_+)$) and
their differentials on $\Gamma$ are linearly independent.

Moreover, we can assume that the inequality
$$
J=\frac{\mathcal{D}(S_1,\dots,S_n)}{\mathcal{D}(t,x_1,\dots,x_{n-1})}\ne0
$$
holds. This inequality is an analog of $S_t\ne0$ at the beginning of
this section and allows us to write $\omega$ in the form
$$
\omega=J^{-1}dx_k\cdot\dots\cdot dx_n.
$$

The integral identity, an analog of (\ref{2.5}), has the form
$$
\int^\infty_0\int_{\mathbb{R}^n}
\big(R\zeta_t+(uR,\nabla\zeta)+aR\zeta\big)\,dx dt +
\int_{\Gamma}e\bigg((v,\nabla)+\frac{\partial}{\partial t}\bigg)
\zeta(x,t)\omega =0.
$$
Integrating the latter relation by parts, we obtain equations for
determining the functions $e$ and $R$ similarly to (\ref{1.3}).

Now we assume that the singular support of the velocity field
is the stratified manifold $\bigcup \Gamma_{i}$
with smooth strata $\Gamma_{t}$ of codimensions $n_i\geq1$.

We also assume that the velocity field trajectories
are transversal to $\bigcup\Gamma$ and are entering trajectories.

Then the general solution of  Eq.~(\ref{0.1}) has the form
\begin{align} \label{a}
\rho(x,t)=R(x,t)+\sum e_i \delta(\Gamma_i),
\end{align}

where $R(x,t)$ is a function smooth outside $\bigcup \Gamma_{i}$,
$e_i$ are functions defined on the strata $\Gamma_{i}$, and the sum
is taken over all strata $\Gamma_{i}$.

The integral identities determining such a generalized solution
have the form
\begin{align} \label{2.6}
&\int^\infty_0\!\int_{\mathbb{R}^n}
(R\zeta_t+(uR,\nabla\zeta)+aR\zeta)\,dx dt
\nonumber\\
&\qquad
+\sum_{i}\int_{\Gamma_{i}} e_i
\bigg[\bigg((v_i,\nabla)+\frac{\partial}{\partial t}\bigg)\zeta(x,t)\bigg]\omega_i=0.
\end{align}
This implies that, outside $\bigcup \Gamma_{i}$,
the function $R$ satisfies the continuity equation
$$
R_t + (\nabla, uR)+aR=0,
$$
and, on the strata $\Gamma_{j}$ for $n_j=1$,
equations of the form (\ref{1.3}) hold, which contain the values of $R$
brought to $\Gamma_{t}$ along the trajectories.
For $n_l=n-k$, $k>1$, on the strata $\Gamma_{l}$,
we have the equations
\begin{equation}\label{2.7}
\frac{\partial}{\partial t} e_l\mu_l +(\nabla,v_le_l\mu_l)=F_l\mu_l,
\end{equation}
where $\mu_l$ is the density of the measure $\omega_l$ with respect
to the measure on $\Gamma_{l}$ which is left-invariant with respect
to the field $\frac{\partial}{\partial t}+\langle
v_l,\nabla\rangle$, and $F_l$ is defined by the following
construction. Denote a $\varepsilon$-neighborhood of $\Gamma_{l}$
by $\Gamma_{l}^\varepsilon$ and denote its boundary by
${\partial}\Gamma_{l}^\varepsilon$. Let us consider the integral
appearing after integration by parts:
$$
\int_{{\partial}\Gamma_{l}^\varepsilon}\zeta\rho u_{nl}
 \omega^{\varepsilon}_l,
$$
where  $u_{nl}$ is the normal component of velocity $u$ on
${\partial}\Gamma_{l}^\varepsilon$,  $\omega^{\varepsilon}_l$ is
the Leray  measure on ${\partial}\Gamma_{l}^\varepsilon$, $\zeta$
is a test function. Passing to the limit as $\varepsilon\to 0$ we
obtain

$$
\lim_{\varepsilon\to0}\int_{{\partial}\Gamma_{l}^\varepsilon}\zeta\rho
u_{nl}\omega^{\varepsilon}_l =\int_{\Gamma_{l}}\zeta\ F_{l}
\omega_l.
$$

It is well known that outside $\bigcup\Gamma_{i}$ the function
$R(x,t)$ can be calculated using the famous Cauchy formula
\begin{equation}\label{3.8}
R(x.t)=\rho_{0}(x,t)\big|\frac
{Dx}{Dx_{0}}\big|^{-1}\exp(-\int^t_0adt')
\end{equation}
where $\rho_{0}$ ia a constant along the trajectories of the field u
outside  $\bigcup\Gamma_{i}$, $\big|\frac {Dx}{Dx_{0}}\big|$ is the
jacobian of the mapping corresponding to the shift along the
trajectories of $u$ and the integral under exponent is calculating
along the trajectories of the field $u$.

This formula implies that the limit as $\varepsilon\to 0$ of the
above integral exists.

 We note that it follows from the above that the
function $R$ is determined independently of the values of $v_i$ on
the strata under the condition that the field trajectories enter
$\bigcup\Gamma_{i}$.

In conclusion, we consider the case where the coefficient $a$ has
a singular support on $\bigcup\Gamma_{i}$, i.e.,
$$
a=f(u).
$$
In this case, we set
$$
a\rho=\check{a}\rho+\sum f(v_i) e_i \delta(\Gamma_{i}).
$$
where $\check{a}=f(u)$ outside $\bigcup\Gamma_{i}$. We note that
such a choice of the definition of the term $a\rho$ is not unique in
this case. But, first, it is consistent with the common concept of
measure solutions (see [3],[5]) and, second, it is of no importance
for the construction of the solution outside $\bigcup\Gamma_{i}$
for the case in which the trajectories $u$ enter
$\bigcup\Gamma_{i}$.

In this case, identity (\ref{2.6}) takes the form
\begin{align}\label{2.11}
&\int^\infty_0\int_{\mathbb{R}^n}
\big(R\zeta_t+( uR,\nabla\zeta)+f(u)R\zeta\big)\,dx dt
\nonumber\\
&\qquad
+\sum_{i}\int_{\Gamma_{i}}e_i\Big[
\Big((v_i,\nabla)+\frac{\partial}{\partial t}+f(v_i)\Big)\zeta(x,t)
\Big]\omega_i=0,
\end{align}
and Eq.~(1.7) can be rewritten in the form
\begin{equation}\label{2.12}
\frac{\partial}{\partial t}(e_l\mu_l) +( \nabla,v_le_l\mu_l)
+f(v_l)=F_l\mu_l.
\end{equation}

All the afore said gives the following statement.
\begin{theorem}
Let that the following conditions be satisfied
for $t\in[0,T]$, $T>0$:

{\rm(1)} $\bigcup\Gamma_{i}$ is a stratifies manifold with smooth strata $\Gamma_{i}$;

{\rm(2)} the trajectories of the field $u$ are smooth outside $\bigcup\Gamma_{i}$,
enter $\bigcup\Gamma_{i}$ and do not intersect outside $\bigcup\Gamma_{i}$;

{\rm(3)} equations (\ref{2.12}) are solvable on the strata $\Gamma_{i}$;

{\rm(4)} the Kirchhoff laws are satisfied on the intersections of strata $\Gamma_{i}$.

Then there exist a general solution to the continuity equation~(0.1)
with $a=f(u)$ in the sense
of the integral identity (\ref{2.11}).
\end{theorem}

\section{The Maslov tunnel asymptotics}

We recall that the asymptotic solutions of a general Cauchy problem
for an equation with pure imaginary characteristics was first
constructed by V.~P.~Maslov [8]. In the present paper, we consider
only the following Cauchy problem
\begin{equation}\label{t.1}
-h\frac{\partial u}{\partial t}
+P\bigg(\overset{2}{x},-h\overset{1}{\frac{\partial}{\partial
x}}\bigg)u=0, \qquad u(x,t,h)|_{t=0}=e^{-S_0(x)/h}\varphi^0(x),
\end{equation}
where $P(x,\xi)$ is the (smooth) symbol of the Kolmogorov--Feller
operator [9], $S_0\geq0$ is a smooth function, $\varphi^0\in
C^\infty_0$, $ h\to+0$ is a small parameter characterizing the
frequency and the amplitude of jumps of the Markov stochastic
process having having transition probability given by $P(x,\xi)$.
To be more precise, we can have in the mind the following form of $P(x,\xi)$:
$$
P(x,\xi)=(A(x)\xi,\xi) +V(x)+
\int_{\mathbb{R}^n}\bigg(e^{i (\xi,\nu)}-1\bigg)\mu(x,d\nu),
$$
where $A(x)$ is positive definite smooth matrix and $\mu(x,d\nu)$ is a family of
bounded measures smooth with respect to $x$.
The symbol $P(x,\xi)$ can also depend on $t$, we will be more precise later on.

Locally in $t$, an asymptotic solution of problem (\ref{t.1}) can be
constructed according to the scheme of the WKB method, see [8]: the
solution is constructed in the form
$$
u=e^{-S(x,t)/h}\sum_{i=0}^{\infty}(\varphi_i(x,t)h^{i}
$$
in the sense of asymptotic series. In this case, for the functions
$S(x,t)$ and $\varphi_0(x,t)$
we obtain the following problems:
\begin{equation}\label{t.2}
\frac{\partial S}{\partial t}+P\bigg(x,\frac{\partial S}{\partial
x}\bigg)=0,\qquad S(x,t)|_{t=0}=S_0(x),
\end{equation}
\begin{gather}\label{t.3}
\frac{\partial \varphi_0}{\partial t} +\Big(\nabla_{\xi}
P\bigg(x,\frac{\partial S}{\partial x}\bigg), \nabla \varphi_0\Big)
+\frac{1}{2}\sum_{ij}\frac{\partial^2 P}{\partial \xi_i\partial
\xi_j}
\frac{\partial^2 S}{\partial x_i\partial x_j}\varphi_0=0,\\
\varphi_0(x,t)|_{t=0}=\varphi^0(x),\nonumber
\end{gather}

As is known, the solution of problem (\ref{t.2}) is constructed
using the solutions of the Hamiltonian system assumed to exist and
to be smooth
\begin{equation}\label{t.4}
\dot x=\nabla_\xi P(x,p),\qquad x|_{t=0}=x_0,
\end{equation}
$$
\dot p=-\nabla_x P(x,p),\qquad p|_{t=0}=\nabla S_0(x_0).
$$
This solution is smooth on the support of $\varphi_0(x,t)$ for all $t$
such that the
Jacobian $\bigg|Dx/Dx_0\bigg|\ne0$ for
$x_0\in\operatorname{supp}\varphi^0(x)$. We let $g^t_H$ denote the
translation mapping along the trajectories of the Hamiltonian system
(\ref{t.4}).

We recall that the plot
$$
\Lambda^n_0=\{x=x_0,p=\nabla S_0(x_0)\}
$$
is the initial Lagrangian manifold corresponding to Eq.~(\ref{t.2}),
and $\Lambda^n_t=g^t_h\Lambda^n_0$
is the Lagrangian manifold corresponding to
Eq.~(\ref{t.2}) at time $t$.
Let
$\pi:\Lambda^n_t \to \mathbb{R}^n_x$ be the projection
of $\Lambda^n_t$ on $\mathbb{R}^n_x$,
which is assumed to be proper.
The point $\alpha\in \Lambda^n_t$ is said to be essential if
$$
\hat{S}(\alpha,t)=\min_{\beta\in\pi^{-1}(\alpha)}\hat{S}(\beta,t)
$$
and nonessential otherwise. Here $\hat{S}$ is the action
on $\Lambda^n_t$ determined by the formula
$$
\hat{S}(\beta,t)=\int^t_0 p\,dx-H\,dt,
$$
where the integral is calculated along the trajectories of the system
(\ref{t.4}) the projection of whose origin is $x_0=\beta$. As is
known
$$
S(x,t)=\hat{S}(\pi^{-1}x,t)
$$
at regular points where the projection $\pi$ is bijective.

The global in time asymptotic solution of problem (\ref{t.1}) is
given by the Maslov tunnel canonical operator.

To define this operator, following [8, 10]
we introduce the set of essential points
$\bigcup \gamma_{it}\subset \Lambda^n_t$.
This set is closed because the projection $\pi$ is proper, i.e.
that for all $x$ the set of $p$ such that $(x,p)\in \Lambda^n_t, \pi(x,p)=x$
is finite.

Suppose that the open domains $U_j\subset \Lambda^n_t$ form a
locally finite covering of the set $\bigcup\gamma_{it}$. If the set
$U_j$ consists of regular points, then we set
\begin{equation}\label{t.5}
u_j=e^{-S_j(x,t)/h}\varphi_{0j}(x,t)
\end{equation}
where
$$
\varphi_{0j}(x,t)=\psi_{0j}(x,t)\bigg|\frac{Dx_0}{Dx}\bigg|^{1/2},
$$
where $\psi_{0j}(x,t)$ being the solution of the equation
\begin{equation}\label{t.6}
\frac{\partial\psi_{0j}}{\partial t} +( P_{\xi}(x,\nabla
S_j),\nabla\psi_{0j})
-\frac12\operatorname{tr}\frac{\partial^2P}{\partial x\partial\xi}
(x,\nabla S_j)\psi_{0j}=0.
\end{equation}
that exists and is smooth whenever $\bigg|Dx/Dx_0\bigg|\ne0$. The solution
$u_j$ in the domain containing essential (nonregular) points (at
which $d\pi$ is degenerate) is given in the following way: the
canonical change of variables is performed so that the nonregular
points become regular, then we determine a fragment of the solution
in new coordinates by formula (\ref{t.5}) and return to the old
variables, applying the ``quantum'' inverse canonical transformation
to the solution obtained in the new coordinates.

The Hamiltonian determining this canonical transformation
has the form
$$
H_\sigma=\frac12 \sum^{n}_{i=1}\sigma_k p^2_k,
$$
where $\sigma_1,\dots\sigma_n=\operatorname{const}>0$.

The canonical transformation to the new variables is given by the
translation by the time $-1$ along the trajectories of the
Hamiltonian $H_\sigma$. One can prove (see [8],[10]) that the family of
sets $\sigma$ for which the change of variables takes a regular
point into a nonregular is not empty.

Next, the solution near the essential point is determined by the relation
\begin{equation}\label{t.7}
u_j=e^{\frac1h\hat{H}_\sigma}\tilde{u}_j,
\end{equation}
where $\tilde{u}_j$ is given by formula (\ref{t.5}) in the new variables
and
$$
\hat{H}_\sigma=\frac12\sum^{n}_{k=1}\sigma_k
\bigg(-h\frac{\partial}{\partial x_k}\bigg)^2.
$$

On the intersections of singular (containing singular points) and
nonsingular charts (without singular points), we must
match $S_j$ and $\psi_{0j}$. This can be done by applying the
Laplace method to the integral whose kernel is a fundamental
solution for the operator $-h\frac{\partial}{\partial
t}+\hat{H}_\sigma$. This integral appears if we write the right-hand side of (\ref{t.7}) in detail.
In this case, since the solution is real, the Maslov index which is well-known [8] to appear in
hyperbolic problems does not appear. The complete representation of
the solution of problem (\ref{t.1}) is obtained by summing functions
of the type (\ref{t.5}) and (\ref{t.7}) over all the domains $U_j$,
for more detail, see [8], [10].

The asymptotics thus constructed is justified, i.e., the proximity
between the exact and asymptotic solutions of the Cauchy problem
(\ref{t.1}) is proved [8, 9]. More precisely it is proved that at
the  points of the set $\pi(\bigcup\gamma_{it}$ where the projection
$\pi$ is bijective the following estimate holds:
$$
u(x,t,h)-u_j= O(h)
$$

In the preceding case we noted that values of the solution of the
continuity equation at nonregular points are independent of the
values of the solution on the singularity support (of course, the
inverse influence takes place) by the condition that the velocity
field trajectories enters the singular support.

In the case of the canonical
operator construction briefly described above, the relation between the
solutions at essential and
nonessential point is also unilateral, namely, the essential points
are ``bypassed'' using (\ref{t.7}), but the values of the functions
$\tilde{\psi}_{oj}$ contained in $\tilde{u}_j$ on the singularity
support do not determine the values at the regular points (but the
converse is not true).

Now we note that the function $S(x,t)$  such that
$$
S(x,t)|_{U_j}=S_j(\pi^{-1}(\alpha),t)
$$
is globally determined and continuous at points of the domain
$\pi(\bigcup\gamma_{it})\subset\mathbb{R}^n_x$. We denote this set
by $\bigcup\Gamma_{i}$ and assume that this is a stratified
manifold with smooth strata $\Gamma_{it}$ of different codimensions.
We note that, for example, if the inequality
$\nabla(S_i(x,t)-S_j(x,t))\ne0$ holds while we pass from one branch
$\Lambda^n_t\cap\bigcup\gamma_{it}$ to another, then the set
$\pi\{(\tilde{S}_i-\tilde{S}_j)=0\}$ generates a smooth stratum of
codimension $1$. In the one-dimensional case, all strata are points
or curves on the $(x,t)$-plane (under the above assumptions about the
singularities being discrete).

Now we consider the equation for $\psi^2_{0j}$.
We denote this function by $\rho$ and then obtain
\begin{equation}\label{t.8}
\frac{\partial \rho}{\partial t}
+(\nabla,u\rho)+a\rho=0,
\end{equation}
where $u(x,t)=\nabla_{\xi} P(x,\nabla S)$ and $a=-\operatorname{tr}
\frac{\partial^2 P}{\partial x\partial\xi}(x,\nabla S)$.

If the condition
$$
\operatorname{Hess_{\xi}} P(x,\xi)>0
$$
is satisfied, then it follows from the implicit function theorem
that $\nabla S(x,t)=F(x,u(x,t))$, where $F(x,u)$ is a smooth
function and
$$
a=f(x,u),
$$
where $f(x,z)$ is again a smooth function.

Let us return to the formula (\ref{a}) and denote the regular (in
the sense of distributions) part of $\rho$ by $\rho_{\text{reg}}$.

 Thus, we can formulate  the
following theorem.

\begin{theorem}
Suppose that the following conditions are satisfied
for $t\in[0,T]$, $T>0$:

{\rm(1)} There exists a smooth solution of the Hamiltonian system
{\rm(\ref{t.4})}.

{\rm(2)} The singularities of the velocity field $u=\nabla_\xi
P(x,\nabla S)$ form a stratified manifold with smooth strata and
$\operatorname{Hess_\xi} P(x,\xi)>0$.

{\rm(3)} There exists a generalized solution $\rho$ of the Cauchy problem
for Eq.~{\rm(\ref{t.8})} in the sense of the integral identity {\rm(\ref{2.7})}.

Then at the points of $\pi(\bigcup\gamma_{it})$ where the projection
$\pi$ is bijective, the asymptotic solution of the Cauchy problem
{\rm(\ref{t.1})} has the form
$$
u=\exp(-S(x,t)/h)(\sqrt{\rho_{\text{reg}}}+O(h)).
$$
\end{theorem}

This theorem is a global in time analog of the corresponding
Ma\-de\-lung observation about local solutions of Schroedinger type
equations.

Now we demonstrate the connection between solutions of continuity
and transport equations. It is easy to see that by construction a
transport equation solution is equal to $\rho_{\text{reg}}$ out of
the points of singular support. More that one can define
$\sqrt{\rho}$ globally in distributional sense and thus defined
square root is equal to $\rho_{\text{reg}}$. To prove this statement
it is sufficient to note that $\delta$-shock type solution to
continuity equation in the form (\ref{a}) can be obtained
\cite{16,DS4}) as a weak limit of a weak asymptotic solution of the
form
\begin{align} \label{a1}
\rho(x,t)=\rho_{\text{reg}}+\sum e_i \delta_{\varepsilon}(\Gamma_i),
\end{align}
where $x\in R^n$, ${\varepsilon}$ is an axillary small parameter and
$\delta_{\varepsilon}$ is a regularization of $\delta$-function of
the form
$$
\delta_{\varepsilon}=\varepsilon^{-(n-n_i)}\omega(S_{i}/\varepsilon).
$$
Here $S_i$ are smooth functions such that $\nabla S_i|_{S_i=0}\ne0$
and
 the equation $S_i=0$ defines the strata $\Gamma_i$,
$n_i=\dim \Gamma_i$ and $\omega=\omega(\eta)$ belongs to the
Schwartz space of test functions. Without loss of generality one can
consider the case $n_i=0$ in $(x,t)$ half-plain for  each fixed $t$.
In this case the function $S_i$ can be chosen in the form
$S=x-\phi(t)$ and our statement becomes
\begin{align} \label{a2}
\lim_{\varepsilon \to 0}\int_{R^1}\bigg(
\sqrt{(\rho_{\text{reg}}+e_i\varepsilon^{-1}\omega((x-\phi)/\varepsilon))}-\sqrt{\rho_{\text{reg}}}\Bigg)\varphi(x)dx=0,
\end{align}
 for arbitrary test function $\varphi(x)$ from $C_0^{\infty}$.

  After the change of variables $\eta=(x-\phi)/\varepsilon$ the (\ref{a2})
 gets the equivalent form
\begin{align} \label{a3}
\lim_{\varepsilon \to 0}\sqrt{\varepsilon}\int_{R^1}\bigg(
\sqrt{\varepsilon\rho_{\text{reg}}+e_i\omega(\eta)}-\sqrt{\varepsilon\rho_{\text{reg}}}\Bigg)d\eta=0,
\end{align}
which is obviously true under the trivial inequality $\sqrt{1+z}\le
1+z/2$. Indeed consider the domain in $R^1_{\eta}$ where
$\omega^{\alpha}<\rho \varepsilon$ for some $\alpha, 0<\alpha<1$.
Then the mentioned inequality gives the estimate
$O(\varepsilon^{1/2})$ for the integral under limit in (\ref{a3}).
In the case $\omega^{\alpha}\ge\rho \varepsilon$, because the
function $\omega$ is decreasing (belongs to the Schwartz space), the
measure of corresponding domain can be estimated as
$(\rho\varepsilon)^{-1/N}$ where $N$ is arbitrary positive number.
This completes the proof.

\section{Particular cases}

  The theorem stated in the previous section requires that some assumptions
are satisfied. The most restrictive is the item 3 in the theorem
above-that is the existence of the global generalized solution to
continuity equation. Under the above made assumptions it is possible
to construct this solution using characteristics, but only in the
case where the structure of singular support of $u$ is not changing
in time-all sections of the stratified manifold introduced above by
planes $t=\text{const}$ are diffeomorphic. A more complicate
situation arises, when the singularities of the velocity field
change their structure. In this case the problem of the construction
of a global in time generalized solution to the continuity equation
has not been solved yet. The obstacle is that in this case usually
one has no global in time expression for the velocity field $u$. In
turn this does not allow to apply formula (\ref{3.8}) to construct
global solution to the continuity equation. In multy-dimensional
case as far as we know there is only one result concerning to shock
wave generation [14] which allows to construct global in time
approximation of the shock wave formation process. But this is
slight different from the construction that we needs here. In the
one dimensional case the situation is better and we have all needed
formulas.

We begin with the spatially homogeneous case. Here the problem is
equivalent to the one of constructing a formula for a global
solution to conservation law equation

\begin{equation}\label{d.1*}
\frac{\partial v}{\partial t}+\frac{\partial P(v,t)}{\partial x}=0
\end{equation}

Here $P(-h\frac{\partial}{\partial x},t)$ is the same operator as in
(\ref{t.1}) but assumed to be independent of x with the symbol $P(\xi,t)$ and
$v={\partial S}/{\partial x}$. The velocity field $u$  in this case
is $P_{\xi}(v,t)$. In [11] a construction of the
global solution to the continuity equation where the velocity field
is given by the solution of the equation (\ref{d.1*}) was given. Because the
set of singular points is discrete by our assumptions, without loss
of generality one can consider the case were only one point of
singularity appears. Denote the corresponding (smooth) initial
condition by $u_0$, the instant where the singularity appears by $t^*$
and the point of singularity by $x^*$.

\textbf{The first step} of construction suggested in [11], [12]
(see also \cite{D1}) is that we change $u_0$
in a small neighborhood of of origin $x_0^*$ of the
trajectory coming to $x^*$ when $t=t^*$. We denote this new part of
initial data as $u_1(x_0)$ for $x_0\in(x_0^*-\beta, x_0^*+\beta)$,
$\beta\to0$ and assume

\begin{equation}\label{d.5*}
\varepsilon\beta^{-1}\to0, \qquad \varepsilon\to0.
\end{equation}

We define the function $u_1=u_1(x_0,t)$ as a solution of implicit
equation
\begin{equation}\label{d.11*}
P'_\xi(u_1,t)=- K(t)x_0+b(t),
\end{equation}
The latter equation is solvable under the condition
$\operatorname{Hess_\xi} P(x,\xi)>0$, formulated above.

The functions $K(t)$ and $b(t)$ are defined from the condition of continuity
of the characteristics flow, i.e
$$
u_1(x_0^*-\beta,t)=u_0(x_0^*-\beta,t),\qquad
u_1(x_0^*+\beta,t)=u_0(x_0^*+\beta,t)
$$

 It is easy
to check that this choice of $u_1$ provides that the Jacobian
$\big|Dx/Dx_0\big|$ is identically equal to $0$ for $t=t^*$ and
$x_0\in(x_0^*-\beta, x_0^*+\beta)$. Here we remove from usual
topological concept of general position considering the situation of
identical equality that can be destroyed by small perturbation. But
this construction follows from the algebraic concept  and allows to
present the solution of (\ref{d.1*}) in the form of linear
combination of Heaviside functions (see [11]).

\textbf{The second step} of our construction of an approximation is a
modification of the definition of characteristics. We set
\begin{equation}\label{d.2*}
\dot x=(1-B)P'_{\xi}(u_1(x_0,t ),t)+Bc,\qquad
x_0\in(x_0^*-\beta,x_0^*+\beta)
\end{equation}

and

$$
\dot x=P'_{\xi}(u_0,t),
$$

where $x_0$ does not belong to  $(x_0^*-\beta,x_0^*+\beta)$,
$$
c=\frac{P(v(x(x_0^*+\beta,t),t))
-P(v(x(x_0^*-\beta,t),t))}{v(x(x_0^*+\beta,t),t)-v(x(x_0^*-\beta,t),t)}
$$
Initial data for (\ref{d.2*}) are the following:
$$
x\bigg|_{t=0}=x_0+A\varepsilon,\qquad \varepsilon>0.
$$
The function B in (\ref{d.2*}) has the form
$B=B((t-t^*)/\varepsilon)$ and B(z) is smooth, monotone and
increasing from 0 to 1 for $z\in(-\infty,\infty)$. Similarly to
[11], [12] one can prove that there exist an
$A=\operatorname{const}$ such that the Jacobian $\big|Dx/Dx_0\big|$
calculated using the above introduced characteristics is not equal
to zero, but it is of order $O(\varepsilon)$ when
$t\ge{t^*+O(\beta)}$ when $x_0\in(x_0^*-\beta,x_0^*+\beta)$. Using
the velocity field generated by $\dot x$ we can construct global in
time (smooth) solution of the continuity equation in the form
(\ref{3.8}). After that, passing to the limit as $\varepsilon\to0$
we will obtain the generalized solution of the continuity equation in
the sense of definition from Sec.1 just like it was done in [12].

\textbf{Spatially inhomogeneous one dimensional case}.

We will follow the scheme introduced above. The case under
consideration can be treated in the same way as the previous one
with modifications. Firstly, we will assume that the symbol
$P=P(x,\xi)$ does not depend on t. In this case this assumption (which means
that the mapping $g^t_P$ is invertible)will be
used to construct the insertion to initial data. In the previous
case we did it using the implicit function theorem, see (\ref{d.11*}).

Let $\Lambda^1_0$ be a smooth nonsingular (w.r.t the projection $\pi$)
curve in the $(x,p)$ space, which is a Lagrangian
manifold corresponding to initial data for our problem. We consider
Lagrangian manifold $\Lambda^1_{t^*}=g^{t^*}_P\Lambda^1_0$  and
assume that there is only one point singular with respect to
projection on $x$-axis and its projection is $x^*$. Let $\beta$ be
the same as above. Let us set $t_1^*=t^*+\beta$. Because of
the assumption that $P''_{\xi\xi}$ is positive, we have that for
$t=t_1^*$ the Lagrangian manifold $\Lambda^1_{t^*_1}$ has two parts
which contain essential points and these parts form a shock wave
type curve with the jump at the point $x^*_1$ where
$S_{left}(x^*_1,t^*_1)=S_{right}(x^*_1,t^*_1)$. We connect these
parts by a vertical line and thus obtain a new Lagrangian manifold,
which is piecewise smooth continuous curve with two angle points
(ends of the vertical part, the distance between them of order
$\beta$). We denote this manifold by $\hat\Lambda^1_{t^*_1}$ and
apply the mapping $g_P^{-t_1}$ for sufficiently small $t_1$ to this
manifold. This mapping obviously exists  and is a diffeomorpfism
because our Hamiltonian $P$ does not depend on $t$. We consider the
obtained manifold $g_P^{-t_1}\hat\Lambda^1_{t^*_1}$ as the new
Lagrangian manifold corresponding to our problem for
$t=t^*_1-t_1$changing the manifold $\Lambda^1_{t^*_1-t_1}$ by
$g_P^{-t_1}\hat\Lambda^1_{t^*_1}$. As it was said above the latter
manifold is piecewise smooth curve with two angle points and all
points of the curve outside of the part between these angle points
are regular. Moreover there exist a sufficiently small $t_1$ such
that the part of the curve between these angle points contains only
regular points-these statements are the consequence of the
positivity of $P''$,  its stationarity and the possibility to choose
$t_1$ small enough (and independent on $\varepsilon$).

Denote the projections of the mentioned above angle points on the
manifold $g_P^{-t_1}\hat\Lambda^1_{t^*_1}$ to the $x$-axis by $a_1< a_2$
and note that $\big|a_1-a_2\big|$ is of order $\beta$.

Like in the previous example we introduce the new characteristics
system
\begin{equation}\label{d.3*}
\dot x=(1-B)P'_{\xi}(x(x_0,t ),p(x_0,t))+Bc,
\end{equation}
$$
\dot p=-(1-B)P'_x(x(x_0,t ),p(x_0,t),\qquad x_0\in(a_1,a_2),
$$

and

\begin{equation}\label{d.7*}
\dot x=P'_{\xi}(x,p),
\end{equation}
$$
\dot p=-P'_x(x,p)
$$
when $x_0$ does not belong to  $(a_1,a_2)$. We have set
\begin{equation}\label{d.8*}
c=\frac{P(v(x(a_2,t),p(a_2,t)))-P(x(a_1,t),p(a_1,t))}{p(x(a_2,t),t)-p(x(a_1,t),t)}
\end{equation}
Initial data for (\ref{d.3*}), (\ref{d.7*}) are the following:

$$
x \big|_{t=0}=x_0+A\varepsilon,
$$
$$
p \big|_{t=0}=p_0(x_0),
$$
where $(x_0,p_0(x_0))$=$g_P^{-t_1}\hat\Lambda^1_{t^*_1}$. The expression
in the right hand side of
(\ref{d.8*}) is the direct analog of the well known Rankine-Hugoniot
expression for the velocity of the shock propagation. In the case under
consideration it is the velocity of the point $\check x$ on
$x$-axis, where $S_{left}(\check x,t)=S_{right}(\check x,t)$.

By the assumption we have only one singular point  if we are
considering the family of manifolds $\Lambda^1_t$, $t\in[0,t^*]$. We
also have by construction that the jacobian $J= Dx/Dx_0$ calculated
using  the solutions of the system (\ref{d.3*}) is not equal to zero.
More precisely we have
$$
\lim_{\varepsilon\to0} J=H(t^*-t)J_0,
$$
where $J_0$ is the Jacobian calculated using the solutions of
(\ref{d.3*}) for $B=0$ ($J_0=0$ when $t=t^*$ by construction) and
$$
J \ge H(t^*-t)J_0+C\varepsilon,
$$
where $C=\operatorname{const}>0$. This statement directly follows
from (\ref{d.3*}) if we take the properties of the function $B$ into
account. It means that the velocity field, generated by projections
of the solution of the system (\ref{d.3*}), (\ref{d.7*}) on the $x$-axis
has nonintersecting trajectories for $\varepsilon>0$. Thus we can
use it to construct solutions of the continuity equation. It remains to
note that just like in [12] it is easy co check the the limits of
these solutions will satisfy to the integral identities introduced
in Sec. 1 as the definition of generalized solutions to continuity
equation.

\end{document}